\documentclass[10pt,journal,compsoc]{IEEEtran}
\IEEEoverridecommandlockouts

\usepackage{amsmath,amssymb,amsfonts}
\usepackage{algorithmic}
\usepackage{graphicx}
\usepackage{textcomp}
\usepackage[dvipsnames]{xcolor}
\usepackage{pdfpages}
\usepackage{acronym}
\usepackage{pifont}
\usepackage{pgfplots}
\usepackage{booktabs}
\usepackage[binary-units, per-mode=symbol, detect-weight=true, detect-family=true]{siunitx}
\usepackage{wrapfig}
\usepackage{ifthen}
\usepackage{makecell}
\usepackage[hyphens]{url}
\usepackage{hyperref}
\usepackage{tikz}
\hypersetup{
    colorlinks=true,
    urlcolor={MidnightBlue},
    citecolor={RoyalPurple},
    linkcolor={RoyalPurple},
}

\ifCLASSOPTIONcompsoc
  \usepackage[nocompress]{cite}
\else
  \usepackage{cite}
\fi

\acrodef{WSL}{Windows subsystem for Linux}
\acrodef{PSW}{platform software}
\acrodef{CRT}{common runtime}
\acrodef{SHA-256}{secure hash algorithm 256-bit}
\acrodef{ELF}{extensible linking format}
\acrodef{TCC}{tiny C compiler}
\acrodef{SDK}{software development kit}
\acrodef{PE}{portable executable}
\acrodef{DLL}{dynamic link library}
\acrodef{SMT}{simultaneous multithreading}
\acrodef{ROP}{return-oriented programming}
\acrodef{TSX}{transactional synchronization extensions}
\acrodef{IAS}{Intel attestation service}
\acrodef{TEE}{trusted execution environment}
\acrodefplural{TEE}{trusted execution environments}
\acrodef{LoC}{line of code}
\acrodefplural{LoC}{lines of code}
\acrodef{TCB}{trusted computing base}
\acrodef{API}{application programming interface}
\acrodef{SGX}{Software guard extensions}
\acrodef{EPC}{enclave page cache}
\acrodef{CDF}{cumulative density function}
\acrodef{OS}{operating system}
\acrodef{CFI}{control-flow integrity}
\acrodef{ISV}{independent software vendor}
\acrodefplural{ISV}{independent software vendors}
\acrodef{CPU}{central processing unit}
\acrodef{EPC}{enclave page cache}
\acrodef{ecall}{enclave call}
\acrodefplural{ecall}{enclave calls}
\acrodef{ocall}{outside call}
\acrodefplural{ocall}{outside calls}
\acrodef{TIB}{thread information block}
\acrodef{TEB}{thread environment block}
\acrodef{PMC}{performance monitoring counters}

\newcommand\copyrighttext{  \scriptsize \textcopyright 2020 IEEE.
    Personal use of this material is permitted.
    Permission from IEEE must be obtained for all other uses,
    in any current or future media, including reprinting/republishing this
    material for advertising or promotional purposes, creating new collective
    works, for resale or redistribution to servers or
    lists, or reuse of any copyrighted component of this work in other works.
    Pre-print version. Published in the IEEE Transactions on Dependable and Secure Computing. For the final version, refer to DOI \href{https://doi.org/10.1109/TDSC.2020.3024562}{10.1109/TDSC.2020.3024562}}
\newcommand\copyrightnotice{\begin{tikzpicture}[remember picture,overlay]
\node[anchor=south,yshift=2pt,fill=yellow!20] at (current page.south) {\fbox{\parbox{\dimexpr\textwidth-\fboxsep-\fboxrule\relax}{\copyrighttext}}};
\end{tikzpicture}}

\begin{document}

\title{Malware in the SGX supply chain:\\ Be careful when signing enclaves!}

\author{\thanks{Vlad Cr\u{a}ciun is with Alexandru Ioan Cuza University of Ia\c{s}i, Romania (UAIC); Pascal Felber is with University of Neuch\^{a}tel, Switzerland; Andrei Mogage and Emanuel Onica are also with UAIC; Rafael Pires is with the Swiss Federal Institute of Technology in Lausanne (EPFL). 
emails: vcraciun@info.uaic.ro, pascal.felber@unine.ch, mogage.andrei.catalin@info.uaic.ro, eonica@info.uaic.ro, rafael.pires@epfl.ch.} 
Vlad Cr\u{a}ciun, Pascal Felber \IEEEmembership{Senior Member, IEEE}, Andrei Mogage, \\ Emanuel Onica \IEEEmembership{Member, IEEE},  Rafael Pires \IEEEmembership{Member, IEEE}}

\IEEEtitleabstractindextext{\makeatletter{}
\begin{abstract}
Malware attacks are a significant part of the new software security threats detected each year.
Intel Software Guard Extensions (\ac{SGX}) are a set of hardware instructions introduced by Intel in their recent lines of processors that are intended to provide a secure execution environment for user-developed applications.
To our knowledge, there was no serious attempt yet to overcome the SGX protection by exploiting the weaknesses in the software supply chain infrastructure, namely at the level of the development, build or signing servers. 
While \ac{SGX} protection does not specifically take into consideration such threats, we show in the current paper that a simple malware attack exploiting a separation between the build and signing processes can have a serious damaging impact, practically nullifying \ac{SGX} integrity protection measures.
We also explore two possible mitigations against the attack, one centralized leveraging \ac{SGX} itself, and one distributed that relies on a smart contract deployed on a blockchain infrastructure. 
Our evaluation shows that both methods are feasible in practice and their added costs are acceptable for the offered protection.
\end{abstract}

}

\maketitle
\copyrightnotice

\IEEEdisplaynotcompsoctitleabstractindextext
\IEEEpeerreviewmaketitle

\ifCLASSOPTIONcompsoc
\IEEEraisesectionheading{\section{Introduction}\label{sec:introduction}}
\else
\section{Introduction}
\label{sec:introduction}
\fi

A software supply chain attack can be informally defined as the act of compromising legitimate software packages during their development or distribution phases. 
The number of such attacks showed a tremendous increase recently. 
A NIST forum presentation~\cite{NISTsupplychainattacks} reported seven significant events in 2017 compared to only four during the previous three years. 
One of the most common attack vectors is injecting malicious malware code~\cite{NISTsupplychainattacks,MBytesBlog} into legitimate software packages during or between development and distribution phases, such as upon building or signing. 
The most prominent example is an infected installation package of the well-known CCleaner~\cite{CCleanerMain} application that included a malware deployed in the vendor's build server~\cite{CCleanerAttack}. 
The altered binary file was downloaded by 2.27 million customers, with potentially serious effects ranging from keystrokes recording to stealing secret credentials from users.

Other recent examples of fairly similar supply chain attacks include an embedded malware in software packages released by NetSarang~\cite{NetSarangAttack}, a company that develops secure connectivity solutions, or corrupted packages injected with malicious code used for updates on M.E.Doc~\cite{MEDocAttack}, a popular accounting application suite in Ukraine.
The focus of our paper lies on the severe implications of a supply chain attack against one of the most recent approaches of preserving confidentiality and integrity of applications: Intel \ac{SGX}.

\ac{SGX}~\cite{Costan:2016} is a set of instruction extensions introduced by Intel in their line of commodity processors since the Skylake generation in 2015. 
\ac{SGX} offers developers the benefit of a \ac{TEE} supported in hardware for critical applications or parts of applications requiring enhanced security levels.
The \ac{TEE} can be used as an isolated space for executing code in \emph{enclave} containers where confidentiality and integrity are assured.
The integrity of a software application that will run in the trusted environment is determined by a \emph{measurement} that uniquely identifies the code and initial state inside the enclave~\cite{Costan:2016}.
This measurement is computed as a hash, included in a \emph{signing material} along with additional enclave metadata, and finally signed.
Based on this value, an \emph{attestation} procedure can be performed whenever a third party wants to check if the correct code is actually running in a SGX-capable machine. 
This check requires that each time the enclave is loaded for execution, the measurement be re-calculated and compared for integrity against the initial value. 
If the enclave is altered at any step of the supply chain, resulting in distinct code or initial data, this integrity check will fail. 

Due to the execution in an isolated secure environment, the \ac{SGX} integrity checking procedure is a very attractive countermeasure against the previously described supply chain attacks.
If a critical software application is loaded within a secure enclave, its modification through malicious code injection could be easily detected during the attestation process. 
Unfortunately, as we show in this paper, an attacker can circumvent the SGX integrity protection using a particular attack methodology that consists of injecting pernicious code between the time of building the software binary and its signing that prepares it for the attestation.

Although, in some light, the attack we show can be regarded as generic, since it is applicable to any software package that is vulnerable to tampering before applying a secure signature, the case we present has particular severe implications on the attempt to protect integrity using \ac{SGX}. 
The attack targets the \ac{SGX} signing process and renders useless its enclave measurement in the way this is initially computed, as well as all its subsequent verifications.
As a result, this means that developers and users cannot blindly trust \ac{SGX} in itself as a way to protect their code and data. 
To counter the attack we explore two mitigation variants. 
The first secures the enclave measurement by atomically binding its generation in a signer process with the enclave compilation phase. 
This method is conditioned by the availability of a specially crafted enclave that is capable to integrate the necessary compiler and signer, and is practically feasible especially for building enclaves of smaller size. 
In a second mitigation variant we consider a distributed approach where a set of servers reach consensus on the correct signing material. 
We propose a simple implementation, which makes use of the integrity and consensus offered implicitly by a blockchain network.
This method does not need integrating any special enclave in the SGX supply chain, which also makes it resilient to zero-day exploits over SGX.
It is also applicable for building enclaves of any size, although it comes with some inherent infrastructure requirements.

In Section~\ref{sec:background} we provide details over the measurement mechanism and the way this is currently used in verifying the integrity of an enclave.
Section~\ref{sec:architecture} describes the actual attack scenario, pointing out current vulnerabilities.
Section~\ref{sec:implementation} provides some practical details regarding the effective attack implementation.
Section~\ref{sec:mitigation} discusses possible ways of mitigating the attack.
In Section~\ref{sec:rw} we provide an overview of related work and we conclude in Section~\ref{sec:conclusion}.

\section{SGX Background}
\label{sec:background}

\ac{SGX} has been available as a \ac{TEE} in Intel processors since the Skylake family. 
It is intended to allow applications to safely handle sensitive data when running within secure \emph{enclaves}.
\ac{SGX} provides memory confidentiality, integrity and freshness assurances even against powerful attackers with physical access and full control of system software, including hypervisors and the \ac{OS}.
The security boundary is the \ac{CPU} die, where data is available in plaintext form. 
Outside it, enclaves' data are always encrypted and their respective digests are kept for integrity and freshness checks.

One \ac{SGX} application is formed by the combination of two logical components: trusted and untrusted.
Untrusted code runs in user mode. It is responsible for asking the operating system to allocate enclave memory and is able to perform \acp{ecall} through a special instruction, but it does not have access to enclave memory pages.
Trusted code, on the other hand, is able to access both its own pages and the ones that belong to the same process running in untrusted mode. 
Since the \ac{OS} is not part of the \ac{TCB}, the enclave is not able to directly perform \emph{system calls}, and it can only execute these through \acp{ocall} by relaying the execution of system calls to untrusted code. 
Since trusted code deals with sensitive data, it must be signed and can be \emph{attested} by local or remote parties before being provisioned with secrets.

The development of applications targeted to run within \ac{SGX} enclaves %
includes a mandatory signing step before the executables are able to be deployed and used in production.
This serves two essential purposes:
\emph{(i)}~the code is uniquely associated to \acp{ISV}, making them recognizable by whom interacts with the enclave and accountable for any consequence originated from their product; and
\emph{(ii)}~whoever communicates with the application can have guarantees that the enclaved endpoint has loaded and is actually running the expected code within a genuine \ac{SGX} platform.

\begin{figure}[t]
    \centering
    \includegraphics{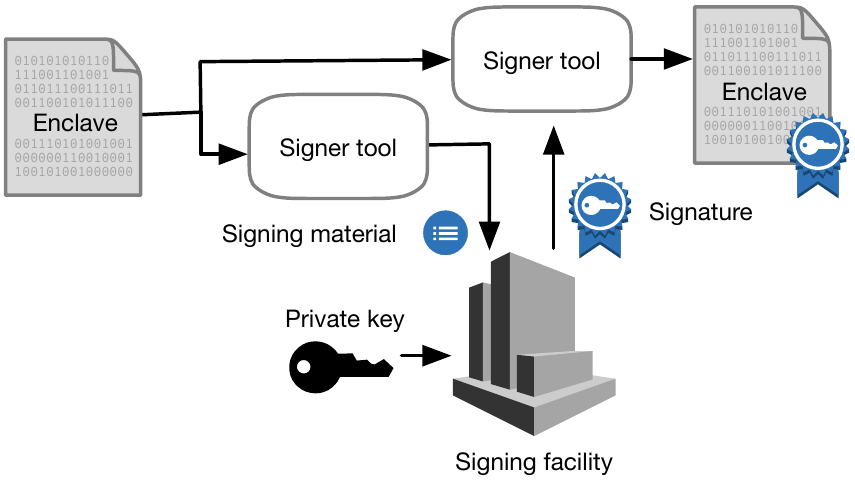}
    \caption{Two-step signing method}
    \label{fig:twostep}
\end{figure}

Intel offers to \acp{ISV} two signing methods~\cite{IntelKeyMngmt}:
\emph{(i)}~single-step method, for development or pre-release modes, which uses a private key locally stored in the building system; and
\emph{(ii)}~two-step method, for release enclaves made by \acp{ISV} who have obtained a \emph{production license} from Intel~\cite{IntelProdLicense}.
With the two-step method, the \acp{ISV} first generate the \emph{signing material}, which is later signed in a different facility that has access to the signing key.
Then, the signature comes back to the building platform and is appended to the enclave's metadata, along with the executable binary.
Figure~\ref{fig:twostep} illustrates the two-step method.

The signing material includes information about the vendor, the date, some attributes, a version number and, especially important to our attack, the enclave \emph{measurement} hash.
This hash corresponds to a digest made upon the enclave's initial state, including data, code and metadata~\cite{Costan:2016}. 
When the enclave is loaded, a hardware implementation of the same procedure performs a \emph{measurement} on the actual content of the running enclave, which has to precisely match the one that was computed during the signing step.
Tamper attempts performed after signing are hence detectable by this protection scheme.
Our attack, however, acts before the signing material is generated and therefore passes undetectable by the \emph{measurement} comparison.

Later on, when interlocutors want to communicate with a running enclave, they should first attest it before sharing sensitive data with it.
Enclave attestation can be performed locally or remotely, the latter being dependent on the former. 
The attestation procedure starts locally through a previously established communication channel, when the attester---which is a platform enclave in case of a remote attestation---sends its identity (\emph{measurement}) to the enclave being attested, or \emph{target}.
This, in turn, calls a special \emph{report} instruction that cryptographically binds the target enclave measurement with other security-related information.
This report's signature can only be checked locally by the attester enclave, as it is generated with a hidden key embedded in the platform and bound to the requester's measurement.
If the platforms' signature is valid, the target's measurement is considered authentic.
In case of remote attestation, another report called \emph{quote} must be generated, which is done by a special enclave provided by Intel within the \ac{PSW} package. This quote, in turn, may be checked by the remote party with the aid of \ac{IAS}.
Since this happens after our attack has been performed, the measurement will correspond to that of the tampered enclave, and therefore it will pass all checks.

\section{Attack Scenario}
\label{sec:architecture}

The purpose of our attack is to corrupt an SGX binary enclave. 
This must happen before the enclave is measured and the resulting measurement is included in the \emph{signing material}.
This way, the rogue enclave will be endorsed by a perfectly valid signature. 
Therefore, we consider a context where an attacker can gain access to a machine where the signing material is generated, as described in Section~\ref{sec:background}. 
Frequently, this is executed on the build server where the enclave is compiled.
We assume that the attacker is able to deploy a malware on such machine.
As referenced in the introduction this is a plausible scenario, multiple similar cases being recently recorded.

The target of our malware is precisely the process that receives the enclave binary as input and generates the \emph{signing material}.
We further refer to this process as the \emph{signer} process. 
The malware will intercept and suspend the \emph{signer} process in order to patch the enclave with malicious code.
Finally, the \emph{signer} process is resumed.
The obtained signing material will include the \emph{measurement} computed over the tampered enclave.
The effective enclave signature will be applied on this altered signing material as nothing abnormal would have happened.
Since the enclave integrity assurance is based on the comparison between the signed measurement and the actual loaded content, any further integrity checks on the maliciously patched enclave will succeed.

\begin{figure}[tb]
	\centering
	\includegraphics{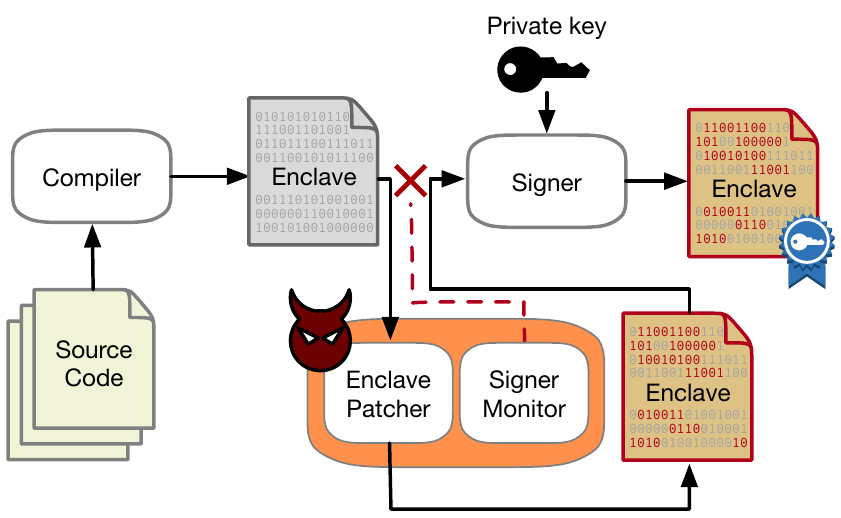}	
	\caption{Attack flow architecture}
	\label{fig:sch1}
\end{figure}

Figure~\ref{fig:sch1} depicts the usual chain for manufacturing an enclave and pinpoints where our malware attacks: after the binary is produced by a compiler and before it is signed.
Note that the figure is representative for the single-step signing method, where the signing material generation and the effective signing are part of the same process, but the flow is similar for the two-step method, with the difference that a separate process performs the effective signing, as shown in Figure~\ref{fig:twostep}. In such case, our attack would strike in the first step, \emph{i.e.}, when the signing material is generated.

\begin{figure}[t]
    	\centering
    	\includegraphics{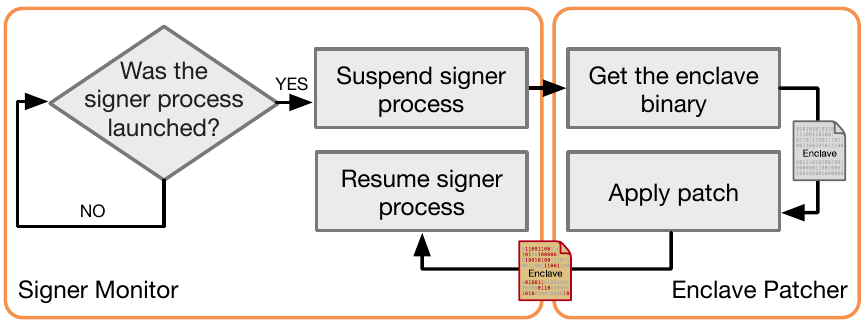}
    	\caption{Malware operation}
    	\label{fig:sch2}
\end{figure}

The malware includes two components used for hijacking and infecting the enclave manufacturing chain: the signer monitor and the enclave patcher. 
The execution flow of the two components is illustrated in Figure~\ref{fig:sch2}.

The \emph{signer monitor} has the role of scanning the running processes until it is able to identify and suspend the signer process.
To achieve that, it uses heuristics such as the process name, input parameters, memory occupancy, hash on certain memory chunks or digital signature.

The \emph{enclave patcher} is composed by malicious code and instructions on where to inject this code.
This depends on the attacker's knowledge about the enclave, ranging from very specific changes on its behavior, like altering some remote server endpoint address, to more generic approaches, like exfiltrating as much information as it can.
We exemplify in the following section a use case where such a malicious patch is hooked to the functions listed in the enclave's \ac{ecall} table, leaks sensitive data out of the enclave and changes input data that is given to the secure enclave space.

\begin{figure}[tb]
	\centering
	\includegraphics{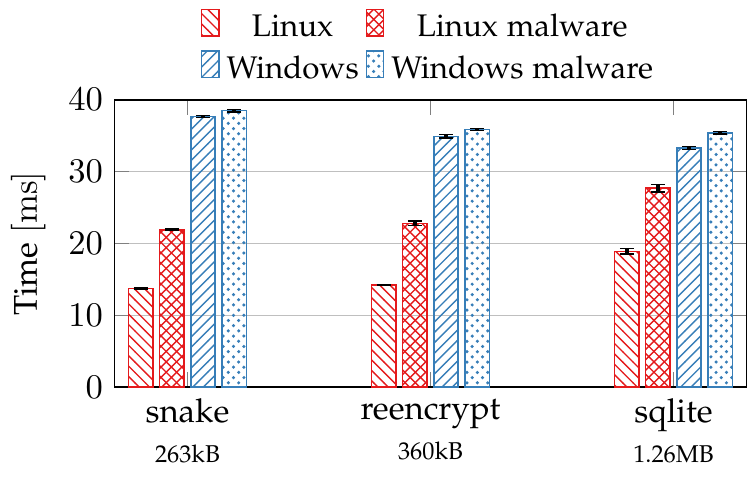}
	\caption{OS comparison between signing times with and without the malware}
	\label{fig:winlinux}
\end{figure}

In order to observe the overhead of our approach, we measured the time it takes for enclave signing with and without activating our malware both in Windows and Linux platforms.
We used three source codes that generated binaries of similar sizes in both platforms.
Results are shown in Figure~\ref{fig:winlinux}.
We measure solely the signer monitor component, as the enclave patcher depends on each specific attack.
All experiments were conducted on a machine with an Intel Core i7-8650U at \SI{1.90}{\giga\hertz}, with \SI{16}{\gibi\byte} of RAM and using Windows 10 Professional x64 build \num{2004}.
Linux measurements were done in the same machine using \ac{WSL}.
Each average accounts for 10 executions and error bars correspond to the 95\% confidence interval.
In general, Windows executions took roughly twice as much time in comparison to Linux ones.
With regard to the malware overhead, we observed on average \SI{8.5}{\milli\second} (Linux) and \SI{1.2}{\milli\second} (Windows) in addition to when the malware is deactivated, which corresponds to increases of \SI{53}{\%} in Linux and \SI{3}{\%} in Windows.

\section{Use Case and Implementation Details}
\label{sec:implementation}

We describe a practical use case about how an attacker could exploit the window of opportunity detailed in Section~\ref{sec:architecture} for getting access to sensitive data and changing it.
Concisely, we inject code to learn about internal data structures and monitor their content in order to modify sensitive data. Besides, we describe a data exfiltration patch.

\begin{figure}[t!]
    	\centering
    	\includegraphics{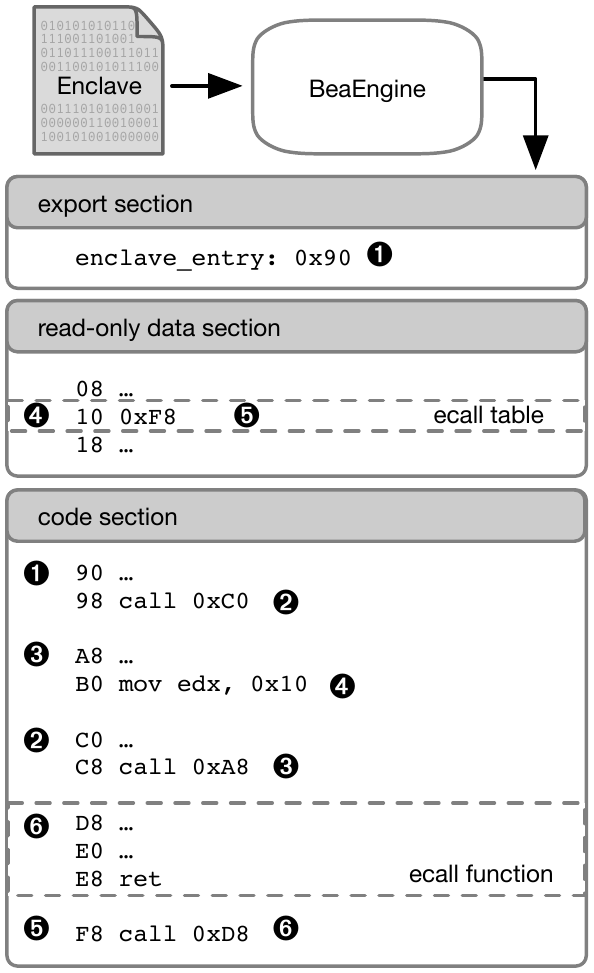}
    	\caption{Steps to find ecall table}
    	\label{fig:findtable}
\end{figure}

\ac{SGX} enclaves are accessed through an instruction called \texttt{EENTER} that transfers the execution to a single entry point in the protected area. 
The specific \ac{ecall} routine address is then fetched from a table to which we refer as \emph{ecall table}.
The enclave patcher finds this table by a series of steps illustrated in Figure~\ref{fig:findtable}.
First, the enclave \ac{DLL} is disassembled with the aid of the BeaEngine library~\cite{BeaEngine}.
It is responsible for parsing and interpreting the \ac{PE} format in which the \ac{DLL} is organized.
The enclave's export data section contains a symbol called \texttt{enclave\_entry}, which is associated to its entry point address (\ding{202}).
By following this address, we find a piece of code that occasionally executes a \texttt{call} instruction to a given address (\ding{203}).
When followed (\ding{204}), this address leads to a chain of other calls to pieces of code that are similar across different enclaves.
Eventually, the ecall table is consulted (\ding{205}). 
In all of our enclave samples, the ecall table was located somewhere in the read-only initialized data section (\texttt{.rdata}) of the \ac{DLL}.
Once we find the table, a similar procedure happens to find the ecall function implemented by the enclave developer (\ding{206} and \ding{207}), since the \ac{SGX} \ac{SDK} adds some wrappers in order to perform security checks before calling the actual enclave code.
All these heuristics were obtained through the analysis of several different enclaves generated by the \ac{SGX} \ac{SDK} version 2.0.101.44281, until a set of patterns allowed us to find the \ac{ecall} table with certainty.
Although this construction may change across different \ac{SDK} versions, the attacker could apply a distinct set of heuristics depending on the version, which is explicitly marked in the \ac{DLL}'s metadata.

\begin{figure*}[t!]
    	\centering
    	\includegraphics{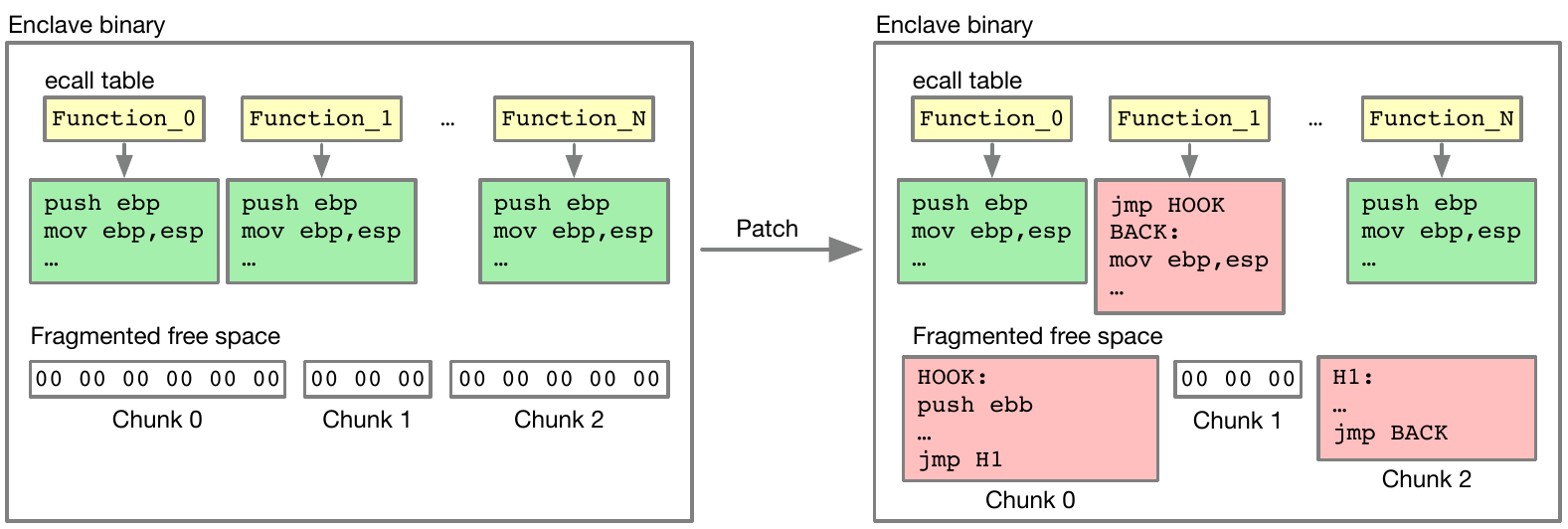}
    	\caption{Enclave patcher}
    	\label{fig:sch3}
\end{figure*}

Once the enclave patcher finds the target \ac{ecall}, it injects a jump instruction in its beginning to a specially crafted piece of code.
We refer to it as the \emph{patch}, whose address is marked with the label \texttt{HOOK}.
Besides the jump instruction to the patch, we add a label (\texttt{BACK}) where the execution continues after executing the hooked code.
Figure~\ref{fig:sch3} shows on the left the initial state of the ecall table and the functions it points to.
In the bottom, we depict a set of disjoint chunks of free memory within executable pages.
These are found by looking for contiguous areas containing only zeros or ones, as these could not possibly refer to any instruction codes.
Such areas are used for placing the patch.
Since there might not be a single chunk big enough for holding the malicious code, the patcher may break it into several pieces linked by jumps and labels.
On the right side of the figure, we illustrate the enclave after applying the patch.
The hooked code is split in two and connected by the label \texttt{H1}.

\begin{figure}[b!]
    	\centering
    	\includegraphics{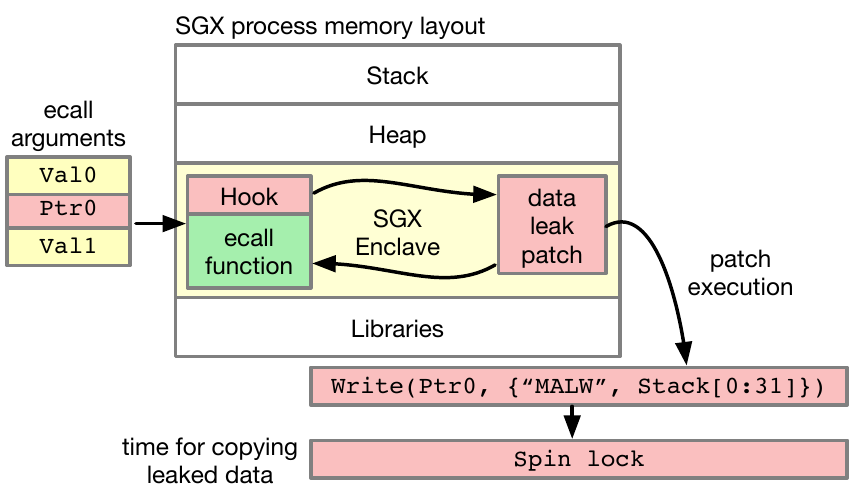}
    	\caption{Exfiltrating enclave data}
    	\label{fig:sch4}
\end{figure}

As for the patch code, we first describe the exfiltration example, depicted as the \emph{data leak patch} in Figure~\ref{fig:sch4}.
The malicious code first tries to identify the arguments of the \ac{ecall} function by looking at the stack.
It checks, among the parameters, if there is an output buffer by evaluating if the  pointer refers to an untrusted piece of memory using auxiliary functions provided by the \ac{SGX} \ac{SDK}\footnote{The SGX SDK provides the functions \texttt{sgx\_is\_within\_enclave} and \texttt{sgx\_is\_outside\_enclave} for this purpose~\cite{intel:2019:sgxsdk}}. 
In the figure, this pointer is referred to as \texttt{Ptr0} and it will be used by the patch to leak sensitive data.
The output buffer \texttt{Ptr0} is shown between two other arguments passed by value: \texttt{Val0} and \texttt{Val1}, whose contents the attacker is interested in leaking along with other local variables found in the stack.

In our prototype, the hook is injected in the beginning of the \ac{ecall}.
One might argue that these values are not interesting for exfiltration at this point, since these would be similar to parameters passed from the untrusted code, which the attacker might already know.
Yet, placing the hook in the end of the function would not be effective, since then the \ac{ecall} would have already written the output buffer and it would not be possible to use it as a leakage vector anymore.
So, the ideal placement is after some computation on local variables has been done, but before the output buffer is written.
For simplicity, we chose to place the hook in the beginning.
We point out, however, that a more useful stack data leakage attack would do it differently.

\begin{figure*}[t!]
	\centering
	\includegraphics{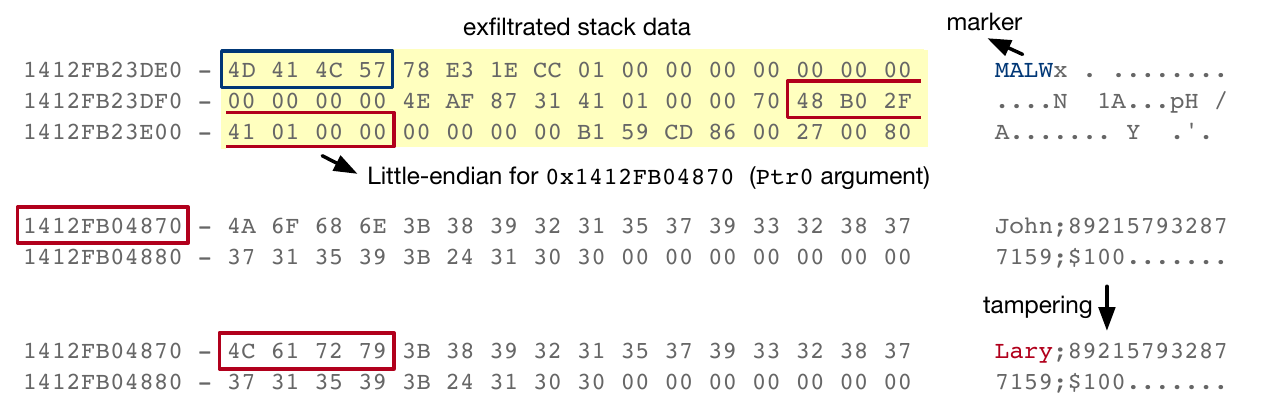}
	\caption{Obtained data}
	\label{fig:sch5}
\end{figure*}

Once the control is diverted to the patch, it copies the stack data to the output buffer.
To confirm that the data leakage has happened and expedite its location, the patch also prepends a marker in the output buffer, referred to as \texttt{MALW} in Figure~\ref{fig:sch4}.
To prevent that the output buffer be overwritten when the \ac{ecall} function is resumed, the patch also uses a spin lock on a boolean variable in the output buffer before it gives back the control to the \ac{ecall}.
Once the untrusted part reads the leaked content, it changes the value of this variable and lets the enclave execution go on.
The leaked data is shown in the first ``hexdump chunk'' of Figure~\ref{fig:sch5}, preceded by the marker.
Although this is an illustrative example, the same techniques could be used for leaking session keys, server credentials or actual payloads decrypted inside the enclave.

Our second experiment, instead of just leaking information, also changes it.
We used the remote attestation end-to-end example~\cite{EndtoEndExample} and the signing tool~\cite{SignerIntel}, both provided by Intel.
It basically performs all the necessary steps for remotely attesting a server and establishing a session key.
Our tampered binaries passed undetected by all attestations, as expected.
We slightly modified the server by adding the transmission of supposedly sensitive information encrypted with the session key, the string \texttt{"John;892157932877159;\$100"} symbolizing, for instance, the destination of some financial transaction.

The enclave patcher, in this case, includes the trampoline to the patch in the end of the decryption function, so that we can modify the information that arrived from the remote server right after it was deciphered with the session key.
The two hexdump chunks of memory on the bottom of Figure~\ref{fig:sch5} show the tampering, by replacing ``John'' for ``Lary''.
Note that we write plaintext decrypted content in the output buffer provided as a parameter of the \ac{ecall} from the former example.
This happens to be untrusted memory area, accessible by the attacker.
In a real world application, any sensitive content must only leave the protected memory in encrypted form.
For the sake of this experiment, however, we used untrusted memory for being able to monitor the tampering when it happened.

In general, our approach resembles other infections of executable files, where malicious code is hooked on the execution flow and the injection is done in the free space within the executable section.
Nevertheless, we identified and provided details on specific features of \ac{SGX} enclaves in our design.
Among these, we locate and analyze the \ac{ecall} table before injecting code in enclave calls, we test whether pointers belong to trusted or untrusted memory areas to locate potential leaking vectors and we synchronize the hooked code with an untrusted agent through spin locks.

\section{Discussion and Mitigations}
\label{sec:mitigation}

Our attack relies on a malware that monitors and intercepts the SGX signer process in order to gain access to the enclave binary.
It will strike where the \emph{signing material} is generated (see Section~\ref{sec:background}).
The attack success depends on altering the enclave binary just \emph{before} this step. 
One could argue that the malware could also intercept the source code sent to the compiler and directly tamper with it.
However, this would be significantly harder. 
It would require the attacker to precisely know the programming language and the particular structure of the code.
Moreover, since such characteristics change for distinct enclave compilations, the attacker should be able to continuously monitor the machine in order to appropriately tune the malware for each attack.
Such capability is beyond our malware scope. 
However, the mitigation we discuss in this section also defends against tampering attacks on both the input as well as the binary output of the compiler. 

An aspect worth mentioning is that the single-step signing method would always expose the signing key to an attacker who has the ability to deploy a malware on the signing platform.
Instead of trying to tamper with the enclave to be signed, the attacker could then just exfiltrate the key.
This would allow the attacker to impersonate the \ac{ISV} by launching rogue enclaves with genuine signatures.
Tampering with the enclave, however, stealthily puts the patch inside the vendor's production code, and potentially gives access to sensitive data more easily.

Since our attack acts just before the signing material is generated, it would succeed even with the supposedly more secure two-step signing method~\cite{IntelKeyMngmt}.
Even though the private signing key is distinctively protected in a different platform, this will be used on the already corrupted signing material that is generated after the tampering takes place. 

With regard to the malware's privileges, we assume in our discussion the worst case when it can escalate to gain super user powers and load kernel modules or manipulate code and data of any process.
A first option to mitigate the attack is a centralized approach that puts compilation and signing inside a dedicated SGX \emph{builder enclave}.
This special enclave must be carefully crafted and accordingly signed by a trusted party.
Ideally, this trusted party would be the compiler vendor, who would offer and maintain the builder enclave.
Sensitive data provisioning (code and key) for building and signing enclaves would be performed through the typical SGX attestation and secret provisioning. 

SGX memory constraints might look like a potential problem for compilation requirements, although such limitation only has impact on compilation time.
The issue can be addressed with appropriate software caching management~\cite{eleos2017} or ordinary memory swapping.
However, this makes the centralized mitigation a better fit for enclaves that are not very large.
We further detail this mitigation variant in Section~\ref{sec:miteval}, by providing a proof-of concept and evaluating its performance costs.

A second option to mitigate the attack would be a distributed approach, where multiple nodes can reach consensus~\cite{bftsmart2014,hybrids2017} on the enclave's hash after each participant compiles its own copy of the source code.
This would not depend on a specific enclave and would be resilient to zero-day attacks targeting SGX.
Also, it would not be constrained by memory usage, at the expense of extra material costs incurred by the distributed infrastructure.
In Section~\ref{sec:distributed} we present a simple prototype relying on a trusted arbitration provided by a blockchain framework.
This would require from the attacker the ability of compromising a certain number of build servers, which is hard to accomplish.

The goal of  both mitigation options is to prevent the attacker from obtaining a signed tampered enclave.
Just exposing the attack is, however, an easier task.
The attack can always be detected if one compares the original untampered enclave with the final tampered and signed one.
When the attacker leaves the original input binary untouched, such comparison can be performed after the signature stage.
For the two-step signing method, the attacker must replace the input with the tampered version, since the signature comes later and it must match with the enclave binary.
As a consequence, it is also possible to detect the attack at the signing material generation step, as long as the original untampered file is preserved.
Comparing the input file before and after launching the signing tool is, however, arguably unusual, and only likely to be performed when being aware of attacks such as ours.
In conclusion, facility of exposure does not nullify the necessity to enhance supply chain security. 

In the following we further explore the two mitigation options just introduced.

\subsection{\label{sec:miteval}Centralized mitigation}

\begin{figure}[tb]
	\centering
	\includegraphics{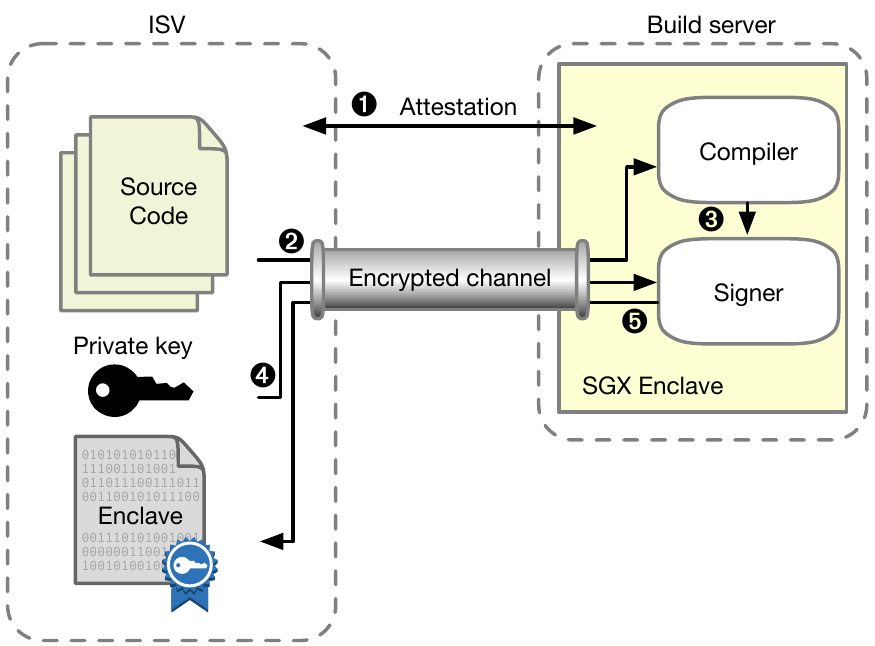}	
	\caption{SGX compilation and signing}
	\label{fig:mitigation}
\end{figure}

We further investigate the centralized mitigation approach described above by designing, implementing and measuring the performance of a combined compiler and signer within SGX enclaves.
Such method would guarantee that the enclave signature corresponds to that of the binary generated by an accredited compiler even if the signing machine is compromised by a super user who has full control of the operating system.
We assume, however, that the source code and private key were safely provided through encrypted channels after an attestation (see Section~\ref{sec:background}) performed by the \ac{ISV}, as illustrated in Figure~\ref{fig:mitigation}.
Once the enclave is attested and the secure communication channel is established (\ding{202}), the \ac{ISV} provides the source code it intends to compile and sign (\ding{203}).
The compiler embedded in the enclave then generates the binary and provides it to the signer (\ding{204}), which also resides in the secure environment.
This, in turn, computes the signature of the generated binary using the private key provisioned by the \ac{ISV} (\ding{205}), to whom it finally sends the final signed enclave (\ding{206}).

\subsubsection{Implementation details}

\begin{figure}[tb]
	\centering
	\includegraphics{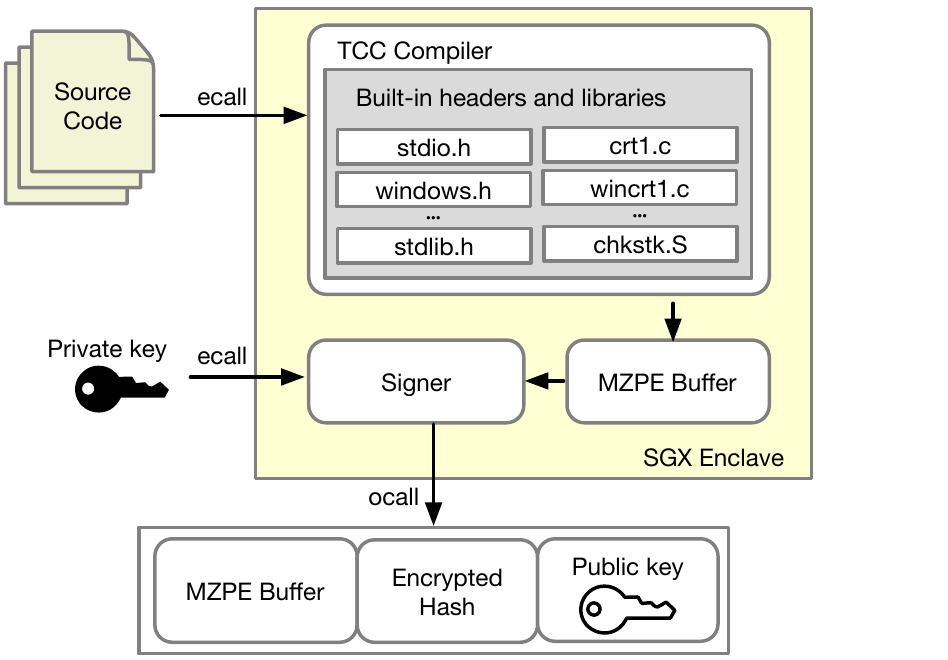}	
	\caption{TCC compiler within an enclave}
	\label{fig:tccsgx}
\end{figure}

We chose the \ac{TCC}~\cite{tinycc} for our experiments.
\ac{TCC} provides support for cross compiling Windows MZPE files (\texttt{.exe} applications and \texttt{.dll} dynamic libraries) and Linux \ac{ELF} files (\texttt{a.out} executables and \texttt{.so} shared objects) for x86, x64 and ARM architectures.
As in our previous experiments, we used Windows x64 as build target.
Figure~\ref{fig:tccsgx} portrays the data flow along with the adaptations we carried out.
In order to support the compilation of legacy code, we had to ship with the enclave some common dependencies, such as standard libraries, headers and the \ac{CRT}. 
In spite of this, the final enclave size, including compiler and signer, accounted for only \SI{2.4}{\mebi\byte}. 

The signer in the enclave receives the compiler output and the signing private key.
We implemented the signature using libraries available in the \ac{SGX} \ac{SDK} (\emph{sgx\_tcrypto} and \emph{sgx\_ippcp}).
Particularly, \ac{SHA-256}, and RSA \num{3072} bits (public exponent equal to \num{3}), as specified by Intel~\cite{intel:2019:sgxsdk}.
The signature is then appended to the enclave binary along with the corresponding public key. This bundle is then sent back to the \ac{ISV}.

We note that the \ac{TCC} compiler does not provide native support for building SGX enclaves. 
We have tuned its environment for building several specific enclaves, as a simple prototype of the centralized mitigation.
\ac{TCC} was conceived with a focus on very low resource consumption, but it does not provide most optimizations offered by other compilers, which typically results in less efficient binaries~\cite{TCCeval}.
We decided to use this compiler due to its simplicity and ease of modifications required to integrate it within an SGX enclave. 
We are aware of its limitations, although we believe that it achieves the purpose of demonstrating our idea.
A dedicated \emph{builder enclave} solution that would natively integrate a full SGX compilation stack should ideally be supported and offered to software developers by Intel in collaboration with the major compiler vendors.

\subsubsection{Evaluation}

\begin{figure}[tb]
	\centering
	\includegraphics{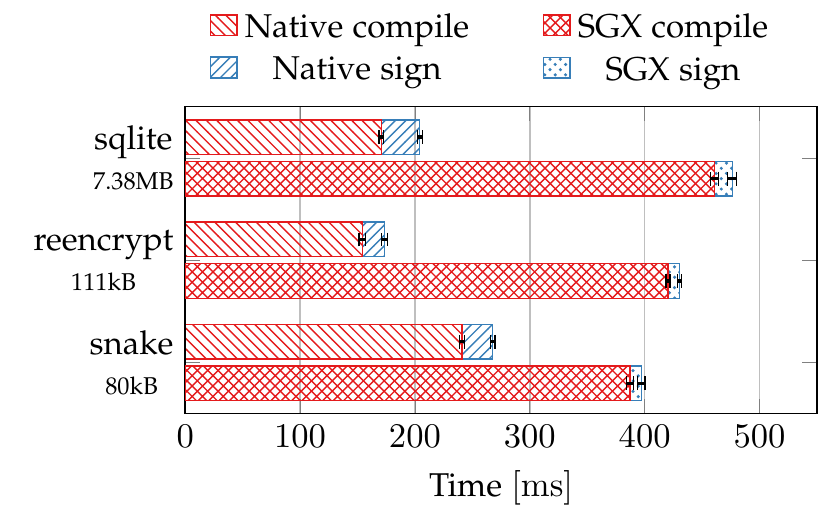}
	\caption{Compilation and signing duration when varying source code sizes}
	\label{fig:inittime}
\end{figure}

The system setup for measuring the compilation and signature durations is similar to the one referred in Section~\ref{sec:architecture}, a machine equipped with an Intel processor i7-8650U at \SI{1.90}{\giga\hertz}, with \SI{16}{\gibi\byte} of RAM and using Windows 10 Professional x64 build \num{2004}.
We tested our mitigation approach on three publicly available SGX enclaves: 
a Snake game version with SGX support \cite{encSnake}, a proxy reencryption application \cite{encreencrypt} and an SQLite database integrated in an SGX enclave \cite{encSQLite}.
Table~\ref{tab:sources} provides information on these samples.
We compiled and signed the corresponding binaries both within a secured SGX enclave using the setup described above, as well as using the native enclave building routine. 
Results are shown in Figure~\ref{fig:inittime}.
Each experiment was repeated \num{10} times and averaged. Error bars correspond to the \SI{95}{\%} confidence interval.

We can notice that the compilation times take longer inside the enclave. 
This is mostly because our builder enclave compilation setup needs a preliminary prepare phase for loading the source files in memory buffers and arranging dependencies.
However, the signing times are slightly lower in the builder enclave. 
The main reason resides in a significant difference between our inside enclave implementation and the outside one. 
The outside version writes the binary on disk and reads it back for the signing step.
Our SGX version, on the other hand, takes advantage of a memory buffer. 
Nevertheless, if the output enclave binary is larger than the stack size after signing in the builder enclave, this might result in a small penalty caused by getting the output out of the enclave in chunks.

In larger projects, we can assume that the whole \SI{93.5}{\mebi\byte} of usable \ac{EPC} memory~\cite{vaucher:2018:sched} could be exhausted. 
In such case we expect the building time to degrade due to memory paging~\cite{pires:2016:scbr} or due to the need of writing temporary files on disk. 
Nevertheless, we did not observe such effects even with our largest samples.
Also, the trade-off between compilation time and security would arguably favor the latter.
Finally, the alternative distributed mitigation approach would be a viable option if such degradation occurs.

\begin{table}[tb]
	\centering
	\caption{\label{tab:sources}Enclave samples used in the benchmark}
	\begin{tabular}{@{}lccc@{}}
	\toprule
	Program   & Code Size & Binary Size & Build Memory Usage              \\ \midrule 
	Micro Snake \cite{encSnake}    & 80kB & 263kB & 2.03MB                           \\
	SGX-reencrypt \cite{encreencrypt}      & 111kB & 360kB & 3.65MB                \\
	SGX-SQLite \cite{encSQLite}    & 7.38MB & 1.26MB & 14.6MB   
\end{tabular}
\end{table}

\subsection{\label{sec:distributed} Distributed mitigation}

In our distributed mitigation prototype, the ISV will use a set of nodes to redundantly and independently build the enclave and to reach consensus on the produced binary.
These nodes can be, for instance, virtual machines deployed by the ISV on a public cloud. 
Each builder node will have an account address on a blockchain network (e.g., Ethereum) allowing it to sign and initiate transactions, namely the possibility to invoke functions in a smart contract. 
Using a blockchain network facilitates both obtaining a trusted consensus and providing integrity guarantees on the signing material. This is due to natural traits offered by blockchain environments, which we briefly introduce next.

\begin{figure*}[t!]
    	\centering
    	\includegraphics[width=0.9\textwidth]{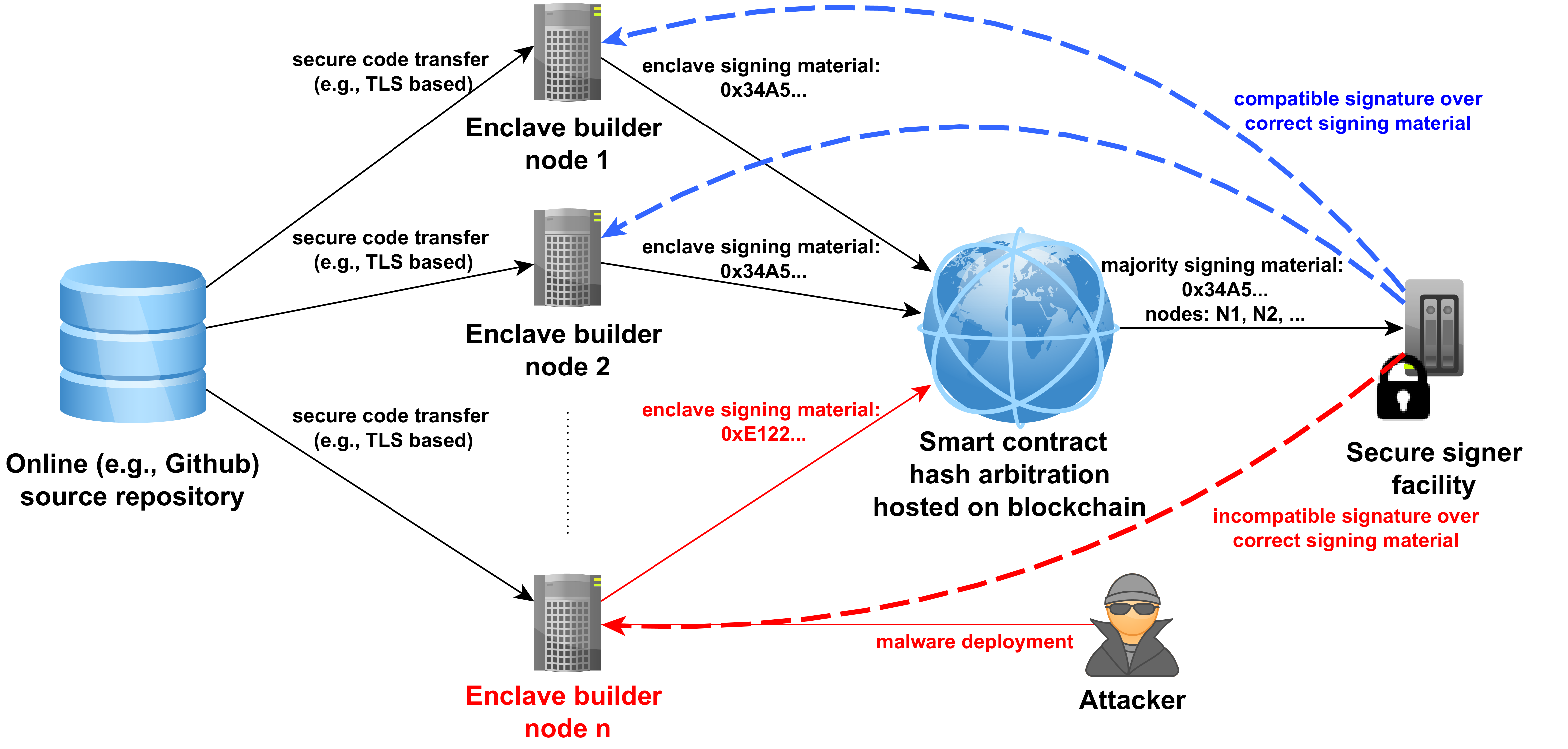}
    	\caption{Distributed mitigation communication flow}
    	\label{fig:distributed}
\end{figure*}

\subsubsection{Blockchain background}

A blockchain network is composed by nodes that maintain a globally replicated data structure: the blockchain ledger.
This ledger evolves following the processing of transactions submitted by the network's clients. 
Transactions are grouped into blocks, and each block is cryptographically linked by a hash to the previous one.
As a consequence, the ledger's structure consists of an append-only chain of blocks, whose consistency and integrity may be checked. 
The peers mutually agree on each new block through consensus protocols that vary depending on the specific blockchain platform. 
A general trait is that the architecture provides decentralized trust in the blockchain data, which are immutable and have guaranteed integrity by design.

Some of the existing blockchain platforms (e.g., Ethereum~\cite{Ethereum}, Hyperledger Fabric~\cite{Hyperledger}) offer the flexibility of running smart contracts, which are small programs that can be executed in the blockchain nodes.
They allow exposing custom functionality to external clients and provide storage in the blockchain to keep the contract data.
Each invocation of a contract function that changes the contract storage counts as a transaction in the ledger, and therefore provides the same blockhain integrity guarantees.
New blocks, including any contract storage changes, are eventually verified and replicated by all nodes in the blockchain network, so that contract storage corruption is considered infeasible.
Also, each external client invoking a smart contract function that counts as transaction has to sign and authenticate its invocation.
Additionally, these clients typically pay a fee that depends on the execution cost of the function and the storage requirements.
These authentication measures can be used to enforce access control for invoking specific contract functions and to prevent DoS attempts to the nodes where these functions are executed. 

\subsubsection{Implementation details}
The distributed mitigation considers a slightly modified version of the two-step signing method referred in Section~\ref{sec:background}, where the enclave's signing material is signed in a separate facility. 
We keep the same attack model, where the attacker gains control over a builder node and corrupts the enclave binary just before the signing material is generated. 
However, it is essential that the separate signer facility does not fall under the attacker control. 
This is a reasonable assumption, since the main purpose of the two-step signing method is to have the effective signing performed on a secure and trusted host.

Given this context, we define an enclave building and signing process that follows the steps depicted in Figure~\ref{fig:distributed}. 
Each builder node will first fetch the enclave sources from the same online repository (e.g., Github). 
To prevent any interference with the code, this transfer can be done via a secure channel using TLS. 
Each node will build the enclave binary and generate the signing material in the typical fashion. 
Next, instead of sending the signing material directly to the signer facility as in the original two-step signing method, each node will submit the generated signing material to a smart contract hosted on a blockchain via a signed transaction. 
The submitted signing material includes the measurement hash of the built enclave. 

The smart contract implements an arbitration for choosing the measurement hash that has a majority.
If no enclave tampering occurred during the build, all hashes submitted by builder nodes should be consistent. 
Otherwise, as depicted in Figure~\ref{fig:distributed}, some might differ. 
The contract will set the majority value as the correct one, and will expose the corresponding signing material to be retrieved by the secure signer facility.
This will be the only signing material to be finally signed.
The process completes similarly to the two-step signing method, with each builder node retrieving the signature from the secure signer facility and using the signer process in a second step to append it to the enclave. 
Corrupted enclaves will obviously fail verification during loading time, therefore the attack attempt will be denied.

The smart contract and all its associated storage are publicly accessible to all nodes involved, including the ones controlled by attackers.
This means that they can read any data stored by the contract. 
However, they are neither able to change the contract code nor to modify the contract data, as this is only possible by executing the functions exposed by the contract. 
Still, since all data used by the contract is publicly accessible, several additional aspects must be considered in respect to the proposed mitigation. 

The blockchain accounts of builder nodes are needed in the contract data initialization, in order to associate their submitted signing material and also limit their actions (i.e., permit only one signing material submission per node and avoid extra fake accounts attempting to gain a majority).
One may consider that the attacker could try to gain control of multiple builder nodes, having this information public. 
However, a blockchain account address typically preserves some level of anonymity, so the real identity of the builder nodes (i.e., their IP addresses) is not directly exposed by the contract data.
Only the ISV and optionally the secure signer facility need to know the association between the nodes IP addresses and their blockchain account addresses.
Moreover, this association is flexible and can be changed, i.e., the ISV can stop a builder node and start a different one at a different IP using the same blockchain address as the former. 
Finally, a corrupted builder node will be detected as its hash will not match the majority's one.
It can hence be blacklisted for participating in future builds.
In summary, it is reasonable to assume that the attacker will not be able to gain control of a majority of the builder nodes. 

The contract storage will be fully open, even beyond the participating nodes. 
Because of this, data can be limited depending on the desired confidentiality level.
Builder nodes could submit only the enclave measurement instead of the entire signing material, which reveals the vendor's identity.
In this case, the signer facility will retrieve from the contract the set of accounts associated with the majority hash, and fetch the signing material directly from one of these.  
In addition, this reduced amount of data in the contract storage will decrease associated storage fees. 

\subsubsection{Evaluation}
We evaluated our proposed smart contract mitigation using the Ethereum blockchain, which is currently the most popular public platform to offer support for running smart contracts.
There are two aspects to be considered: the fee that an ISV would pay for using such a smart contract and the time required by the blockchain network to confirm the transactions following the interaction with the contract.  

Smart contracts in Ethereum are executed by the Ethereum Virtual Machine (EVM), a runtime environment hosted on participating nodes in the network.
Since the network is public, in order to prevent abuse, each contract transaction generates a material cost that has to be paid by the entity requesting it.
This cost is quantified using an internal Ethereum currency named \emph{gas}, and based on a fixed quantum applied for the various transaction opcodes defined at the EVM level (Appendix G of~\cite{Ethereum}).  

Our smart contract has a simple structure.
The interaction with the contract includes a single type of transaction that is cost significant: the submission of the signing material by each builder node. 
We have considered an epoch-based setting, where each builder node is allowed to send exactly one signing material per epoch.
We programmed the contract such that each submission updates a counter associated with the sent signing material.
A decision on the majority is automatically taken following the last builder's submission within the epoch. 
We note that the transaction associated with first submission is more costly than the rest since it is the first that stores a signing material for that epoch (changing storage space from zero is one of the most expensive operations at the EVM level).
The following transactions within an epoch have a constant cost, with minor variations if submitting a different signing material or when triggering a decision on the majority per epoch. 

We have tested our contract on a local Ethereum network simulator and on the Ropsten network, a test network that closely mimics the main Ethereum network operation.
Table~\ref{tab:tx} presents the results for both the case of submitting the complete signing material and the optimized scenario with only the enclave hash.
The latter reduces the contract storage costs, but requires additional communication for retrieving the signing material from one of the builder nodes.  

\begin{table}[tb]
	\centering
	\caption{\label{tab:tx}Submission transaction cost evaluation}
	\begin{tabular}{@{}lrrr@{}}
	\toprule
	Transaction evaluated      & Local (gas) & Ropsten (gas) & Fiat cost (\textdollar)                              \\ \midrule
	First submission (full)        & 448211       &  427875          &  ~1.23  \\
	Other submission (full) & 83809        &  75473            &  ~0.23   \\
	First submission (hash)      & 91322        & 95450             &  ~0.25   \\
	Other submission (hash) & 42636    &  49764            &   ~0.11   \\ \bottomrule
	\end{tabular}
\end{table}

Last column of Table~\ref{tab:tx} provides an estimation of the fee that an ISV would pay in fiat currency (actual monetary costs).
While the gas amount is fixed in Ethereum, the fiat price per gas unit depends on the market.
This price is set at transaction submission and expressed in the cryptocurrency specific to Ethereum (usually in \emph{Gwei}). 
Variations on this price can impact how fast the transaction is confirmed by the network. 
The values in the table were computed based on the median of price paid per gas unit as recorded between August 2019 and August 2020~\cite{etherscan}, which was 14.8 Gwei. 
To estimate the fiat value we used the median conversion rate to US dollars~(\textdollar) during the same period, namely $186 \times10^{-9}$~\textdollar per Gwei~\cite{coinmarketcap}.
An ISV using 10 builder nodes would pay an approximate \textbf{\textdollar~3.30} for building one enclave.
Adding more builder nodes for higher security would cost about \textbf{\textdollar~0.23} more per node. 
The cost could be lowered to a total of \textdollar~1.25 and \textdollar~0.11 per additional node if only the enclave hash is submitted to the contract. 
We note, however, that the Ethereum market is very volatile, recording gas price variations of more than 900\% only in the last year.
For instance, in our first estimations computed in December 2019 the expected price for an average transaction confirmation time reached a low 3 Gwei per gas unit.
This resulted in less than \textbf{\textdollar~0.50} for building one enclave.

Another aspect we considered is the transaction time, which depends on the confirmation that the block where it is included was actually appended to the ledger.
One block has a maximum gas limit for the transactions it includes. 
This is periodically adjusted in the Ethereum network, but is typically close to 10M gas~\cite{ethstats}, which is comfortably larger than the total gas required by 10 builder nodes.
This means that all transactions could fit in a block.
It might happen, however, that other transactions are grouped together with the submissions. 
In our tests on the Ropsten network, the transactions spanned over at most 3 blocks. 
Since the average confirmation time per block in the main Ethereum network is about 15 seconds~\cite{ethstats}, this results in a range of \textbf{15 to 45 seconds} per consensus round.
This will be the dominant time for executing one epoch, which corresponds to one enclave in our distributed protocol.

We emphasize that our results are obtained on the most popular, accessible and widely used \emph{public} blockchain network that supports smart contracts.
Our experiment shows that our distributed mitigation is already feasible in this context.
The time span of a transaction confirmation could be drastically reduced and the material fees removed if the solution would be implemented on a \emph{permissioned} blockchain, such as Hyperledger Fabric~\cite{Hyperledger}.
This is typically run by smaller number of nodes, in closed enterprise environments.

\section{Related Work}
\label{sec:rw}

There is currently very limited published research specifically addressing malware attacks on SGX, and to our knowledge none addressing the context we refer to in our work. 
The research presented in~\cite{malware_sgx:16} is probably the closest approach to our case.
The situation considered is of a malware code that is encrypted, downloaded, decrypted and executed in an enclave that was previously attested as legitimate.
To initiate these steps, the attacker requires a remote bootstrap program that is used to build the initial enclave, attest it, and facilitate the exchange of keys with the attacker for encrypting the malware code.
The decrypted malware code running inside the enclave can subsequently receive other instructions or input from the attacker. 

Most of the documented attacks that target SGX rely on cache exploits. 
In~\cite{MGX:2017}, the authors assume the inclusion of malware in a malicious enclave co-located with a victim enclave. 
The malware performs a \emph{Prime+Probe} side-channel attack through which it is able to recover RSA keys used in the victim enclave. 
The purpose of hiding the malware inside an enclave is to conceal the malicious code, leveraging the SGX protection features to avoid detection. 
However, the attack does not target effectively infecting or corrupting the enclave where the malware resides, which is our case. 

Another work~\cite{schwarz:2019:practicalsgxmalware} also profits from SGX isolation to stealthily operate by leveraging Intel's \ac{TSX} memory-disclosure primitive. 
They show how to bypass the host application interface and execute arbitrary system calls via \ac{ROP}, without collaboration from untrusted code. 
Different from us, they do not target sensitive data operated by enclaves, but rather hijacking the infected machine.

The authors of~\cite{SGXposure:2017} present another attack on SGX that develops on the Prime+Probe technique of recovering cache information.
In this case the attack isolates the core used by the victim enclave from other processes to minimize the noise in the side channel.
Another improvement of the attack is uninterrupted execution by configuring the interrupt controller to not deliver interrupts to the attack core, which could be used to deflect side channel attacks. 
The attack also relies on Intel \ac{PMC} for monitoring cache evictions and also monitors the frequency in order to not miss victim accesses to the cache.
In~\cite{Gotzfried:2017}, the authors also describe an attack that uses CPU pinning and Intel \ac{PMC} in a Prime+Probe approach.
The attack retrieves cache information that leads to an AES key leak. 

The recent Foreshadow attack~\cite{VanBulck:2018} again targets CPU cache leaks, by exploiting a speculative execution bug.
This consists in an unauthorized memory access in transient out-of-order instructions, which can be used before rollback to retrieve confidential data, in a similar manner to the Meltdown attack~\cite{Lipp2018meltdown}. 
Another recently-published side-channel attack~\cite{bhattacharyya:2019:smt} exploits contention on \ac{SMT} and code-reuse. 

In~\cite{Jang:2017}, the authors describe an attack that violates the enclave integrity with the purpose of triggering a processor lockdown.
The attack relies on the Rowhammer approach for flipping bits in the \ac{EPC} memory region.
This is achieved by executing a code snippet inside the enclave that has to find conflicting row addresses in the same memory bank of the EPC. 
The code snippet is supposed to be executed inside a malicious enclave that will be downloaded on a victim machine.
Our attack scenario opens the possibility to corrupt legitimate enclaves with custom malicious code, which could also be such DoS triggering routines injected during the signing process. 

In~\cite{in-toto}, the authors present \emph{in-toto}, a solution that attempts protecting the integrity of an entire software supply chain. 
This consists in a framework where a software project owner has to define a layout of the chain steps. 
Each of the steps requires trusted involved actors to provide signed attestations of metadata associated with that particular step.
The solution is, however, a general one, not addressing any specificities that might appear at the individual steps, as in our case of SGX enclave signing. 
In particular~\emph{in-toto} focuses exclusively on the detection of the tampering, and not on prevention measures.
We believe that our mitigation proposals could be integrated with such a framework.

Several solutions have been proposed for leveraging blockchain immutability in securing integrity of binary files. 
These typically address protecting the integrity of simple binaries in a supply chain, not specifically of an SGX enclave.
The most close to our distributed mitigation is the idea presented in~\cite{Strengele2019}.
Similar to our approach, a hash digest computed over the binary is stored in a smart contract in the Ethereum blockchain, with the purpose to detect if the binary was tampered. 
However, the considered attack surface is limited to altering the binary only \emph{after} the generation of the correct hash.
This is fundamentally different from our scenario, where the attacker tampers an enclave \emph{before} the generation of the signing material. 
The less powerful attack model permits delegating to an external trusted auditor to verify the digest stored in the blockchain, by comparing it to one computed over the binary distributed to users.
This detects an attack but does not directly prevent it.
Our distributed protocol mitigates the attack by not allowing the proper completion of the two-step signing method for a tampered binary, rendering it unusable.

Other work that suggests blockchain as a trusted entity for attestation of software integrity is~\cite{Jesus2018}. 
The authors choose again Ethereum as a root-of-trust platform, proposing the use of smart contracts in several baseline scenarios for safely bootstrapping a device, tampering detection or updating a value securely by a smart meter.
The targeted use cases address mainly the operation of IoT devices, and proposed interactions with a smart contract are mostly described at conceptual level relying on external attestation protocols or manual inspection.
The work is therefore, orthogonal to our attack context, but as well as ~\cite{Strengele2019} it strengthens the common assumption of using blockchain, and in particular Ethereum, for its integrity guarantees.

\section{Conclusion}
\label{sec:conclusion}

We have presented a novel attack in the area of supply chain malware, with a specific target on protection measures that involve the use of SGX.
We provided a practical use case for our attack methodology, which is able to successfully extract sensitive data from the secure enclave space.
This use case is generic enough to be applied to multiple cases of enclaves. 
A malicious entity who has knowledge of a particular enclave functionality can leverage the attack scenario for more specific attacks that can be even more disabling, e.g., changing some particular behavior of the enclave code. 
The flexibility of the attack scenario, which requires essentially just a window of opportunity between the building and the signing of an enclave, makes it quite problematic.

Fortunately, some basic mitigation mechanisms are relatively easy to enforce, as discussed in Section~\ref{sec:mitigation}.
We have shown that protection via compiling and securely signing the binary within a dedicated SGX enclave, as well as a distributed approach that relies on smart contract operation over blockchain, are both feasible practical options.
Also, we believe some other more advanced mitigation options involving cryptographic techniques, such as verifiable secret sharing, could be explored. 
As future work, we consider evaluating and comparing such mechanisms.

\section*{Acknowledgments}
\setlength{\columnsep}{9pt}
\setlength{\intextsep}{3pt}

Some of the activities that contributed to this work were funded by the \emph{European Union{\textquotesingle}s Horizon 2020 research and innovation programme} under grant agreement No 692178. 

\bibliographystyle{IEEEtran}

\renewcommand{\baselinestretch}{1}
\vfill
\pagebreak

\begin{IEEEbiography}[{\includegraphics[width=1in,height=1.25in,clip,keepaspectratio]{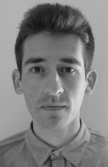}}]{Vlad Cr\u{a}ciun} is a PhD student at the Alexandru Ioan Cuza University of Ia\c{s}i, Romania, studying the field of automated binary analysis. He joined Bitdefender Laboratories in early 2009, being involved in threat cleaning mechanisms and cryptography. His current research interest includes binary instrumentation, symbolic/concolic execution, and control flow analysis.
\end{IEEEbiography}

\begin{IEEEbiography}[{\includegraphics[width=1in,height=1.25in,clip,keepaspectratio]{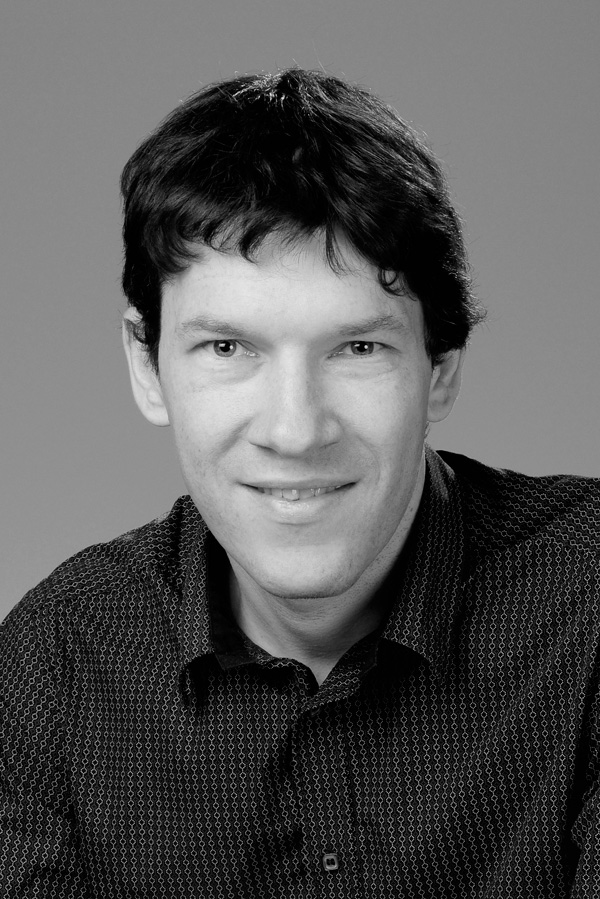}}]{Pascal Felber} received his M.Sc. and Ph.D. degrees in Computer Science from the Swiss Federal Institute of Technology (EPFL). He has then worked at Oracle Corporation and Bell-Labs in the USA, and at Institut EURECOM in France. Since 2004, he is a Professor of Computer Science at the University of Neuch\^atel, Switzerland, working in the field of dependable, distributed, and concurrent systems. He has published over 200 research papers in various journals and conferences. 
\end{IEEEbiography}

\begin{IEEEbiography}[{\includegraphics[width=1in,height=1.25in,clip,keepaspectratio]{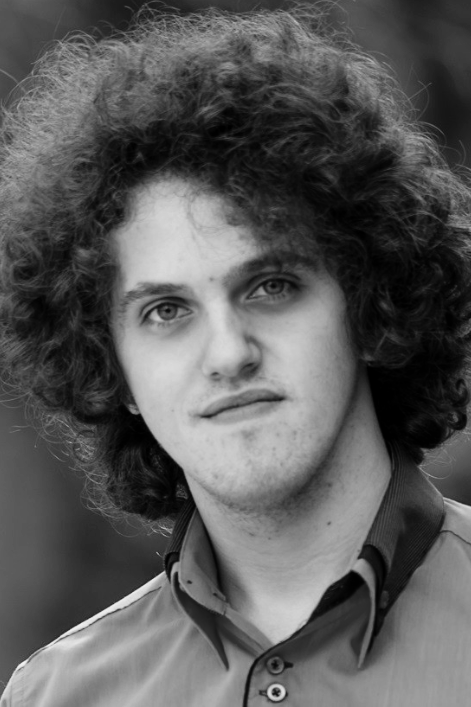}}]{Andrei Mogage} is a PhD student at the Alexandru Ioan Cuza University of Ia\c{s}i, Romania, studying formal methods with a keen interest towards security. He has joined Bitdefender in 2016 and has been involved in cyber threat disinfection and decryption. His current research interests include cryptography, malicious threats and exploitation.
\end{IEEEbiography}

\begin{IEEEbiography}[{\includegraphics[width=1in,height=1.25in,clip,keepaspectratio]{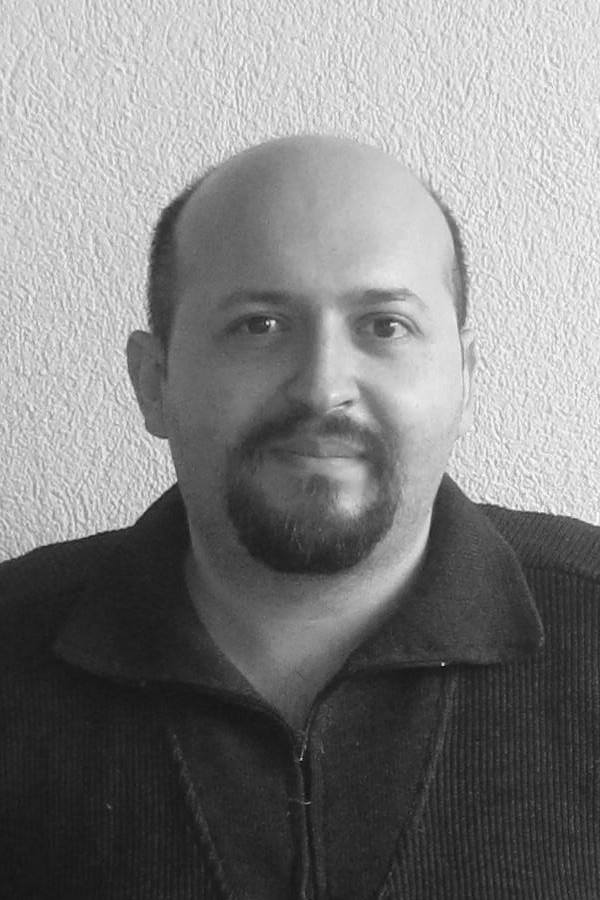}}]{Emanuel Onica} is an associate professor at the Alexandru Ioan Cuza University of Ia\c{s}i, Romania. He received his Ph.D. degree in Computer Science from University of Neuch\^atel, Switzerland in 2014, where he worked as scientific collaborator from 2010 to 2014. 
His current research interests lie in the area of distributed event based systems, where he is the recipient of four awards in international conferences. Most of his work in this field has a focus on privacy and security.
\end{IEEEbiography}

\begin{IEEEbiography}[{\includegraphics[width=1in,height=1.25in,clip,keepaspectratio]{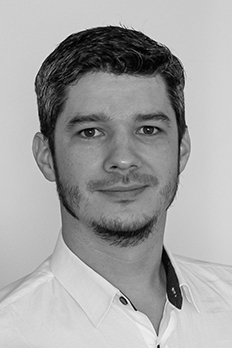}}]{Rafael Pires} is a postdoctoral researcher at the Swiss Federal Institute of Technology in Lausanne (EPFL).
He received his Ph.D. degree in Computer Science from the University of Neuch\^atel, Switzerland, and his B.Sc and M.Sc. degrees from the Federal Universities of Santa Maria and Santa Catarina, respectively, both in Brazil.
His current research interests lie in the usage of trusted execution environments for enhancing security and privacy guarantees in distributed systems.
\end{IEEEbiography}

\vfill

\end{document}